\documentclass[12pt]{article}
\usepackage{amsmath,epsf,amssymb,latexsym,cite}
\usepackage{amsfonts}

\begin{document}

\def\bs{\bigskip}
\def\d{\partial}
\def\Sbu{S_\Sigma}
\def\Sbo{S_{\d \Sigma}}
\def\At{\widetilde{A}}
\def\Ct{\widetilde{C}}
\def\Ch{\widehat{C}}
\def\Ph{\widehat{\psi}_-}
\def\Th{\widehat{T}}
\def\Rt{\widetilde{R}}
\def\Qt{\widetilde{Q}}
\def\Ft{\widetilde{F}}

\def\dpsit{{\delta\psi'}}

\def\intb{\int_{\d\Sigma} dx^0}

\def\Fc{{\cal F}}
\def\Fct{\widetilde{\Fc}}

\def\ep{\epsilon}
\def\epb{\bar{\ep}}
\def\epp{\ep_+}
\def\epm{\ep_-}
\def\epbp{\epb_+}
\def\epbm{\epb_-}
\def\Lc{{\cal L}}
\def\Mc{{\cal M}}
\def\Nc{{\cal N}}

\def\nab{\nabla}
\def\ona{{\overline{\nabla}}}

\def\ov{\overline}
\newcommand*{\dsym}{\mathop{\mathrm{Sym}}\displaylimits}
\newcommand*{\LHS}{\mathop{\mathrm{LHS}}\displaylimits}
\newcommand*{\RHS}{\mathop{\mathrm{RHS}}\displaylimits}
\begin{titlepage}
\begin{center}

\hfill  {Duke-CGTP-03-04} \\
\hfill  {NSF-ITP-03-78} \\
\hfill  {\tt hep-th/0309223} \\
        [22mm]

{\Huge Supersymmetric Boundary Conditions \\ \vspace{2mm} 
for the $\Nc = 2$ Sigma Model}\\ 
\vspace{6mm}

{\Large Ilarion V. Melnikov, M.~Ronen Plesser, and Sven Rinke} \\
\vspace{3mm}
{\small \it Center for Geometry and Theoretical Physics, \\}
{\small \it Duke University, \\}
{\small \it Durham, NC 27708.\\}
\vspace{1mm}
{\small Email : lmel, plesser, rinke@cgtp.duke.edu}\\
\date{\today}

\end{center}
\medskip
\vspace{22mm}

\begin{abstract}
We clarify the discussion of $\Nc =2$ supersymmetric boundary
conditions for the classical $d=2$, $\Nc=(2,2)$ Non-Linear 
Sigma Model on an infinite strip.  Our conclusions about the 
supersymmetric cycles match the results found in the literature.  
However, we find a constraint on the boundary action that is not 
satisfied by many boundary actions used in the literature.          
\end{abstract}

\end{titlepage}

\newpage 

\vspace{5mm}
\section{Introduction}
Weakly coupled Type II string theory compactified on a Calabi-Yau
manifold provides a tractable setting for understanding (or at 
least observing) aspects of stringy geometry \cite{GREENE}.  
Our ability to glean insights into such non-trivial issues is 
largely due to the comparative tractability of the worldsheet 
approach in this compactification.  Although the $d=2$, 
$\Nc =(2,2)$ Non-Linear Sigma Model with Calabi-Yau target space 
is a complicated theory with non-trivial IR dynamics, the existence 
of a well-defined classical limit, the presence of topological 
sectors, and mirror symmetry allow us to draw some rigorous 
conclusions about the theory.  Using these techniques, much has 
been learned about how closed strings probe the geometry.  
Naturally, we would like to learn how objects other than 
fundamental strings probe the geometry. This is, in principle, 
a very difficult task, as it requires a study of complicated 
solitonic objects in the theory.  However, as is well known, it is 
our inexplicable bit of fortune to have access to a large class of 
non-perturbative objects whose fluctuations can be described in 
terms of perturbative degrees of freedom:  namely, D-branes, which 
can be thought of as ``places where open strings can end.''  This 
statement is a bit too naive in general curved backgrounds.  
However, it is the right point of view at the large radius limit, 
where classical analysis is valid. This connection means that 
in the $g_s \rightarrow 0$ limit we can study D-branes by examining 
the open string NLSM.  In this paper we will study the classical 
worldsheet 
theory for an open string whose endpoints are attached to D-branes 
in a Calabi-Yau manifold.  The problem of interest here is to classify
all stable BPS configurations of D-branes.   As a first step, one
could classify the BPS configurations.  In the worldsheet description
a BPS configuration is a set of boundary conditions and a boundary
action preserving an $\Nc = 2$ superconformal worldsheet symmetry with 
integral $U(1)_R$ charges.  What we will do here is to classify all
boundary conditions and boundary actions which classically preserve an
$\Nc = 2$ superconformal symmetry.  
A set of boundary conditions for the NLSM includes a choice 
of a submanifold on which the open string ends, and the BPS conditions
single out particular (minimal) representatives of equivalence classes
under homology (cycles).  A cycle that has a representative
preserving $\Nc = 2$ worldsheet supersymmetry will be called a 
supersymmetric cycle.  

It is simple to extend the NLSM description to include open
strings---one needs to consider worldsheets with boundaries and to
introduce additional background fields that couple to the string
endpoints.  We will work in the $H=d B = 0$ background.  On a
worldsheet with boundaries, the familiar $\Nc = (2,2)$ NLSM can be
modified by adding a local boundary action constrained by boundary
reparametrization invariance and (classical) scale invariance. In
addition, one needs to specify boundary conditions for the fields.
These will be chosen to eliminate the surface term in the variation of
the action.  Physically, this ensures that the bulk equations of
motion continue smoothly to the boundary.  This constraint has
sometimes been called ``locality'' \cite{HIV}.  To preserve an $\Nc =
2$ SUSY we will need to ensure that the boundary conditions and the
action are invariant under the SUSY variations.

Over the years, this analysis has been carried out by several groups 
\cite{OOY,HIV,KO, HASSAN} with results that more or less agree
with an earlier spacetime analysis in \cite{BBS}.  An exception is the
finding in \cite{KO} that special Lagrangian cycles must, in general,
be extended to co-isotropic submanifolds for a complete
classification.  More careful analyses in \cite{ALZ1,ALZ2,LZ}
extended this work to nontrivial spaces.
Our results agree with the consensus, but we clarify several points. 
Namely, we discuss the role of the ``locality'' constraint, the 
necessity for the action {\em and} the boundary conditions to be 
invariant under the preserved supersymmetry, and constraints on the 
boundary couplings.  
Surprisingly, we find that the standard supersymmetric boundary 
coupling that dates back to Callan {\it et al} \cite{CLNY} does not 
satisfy the locality constraint!  Although our analysis does not 
uniquely determine this the boundary coupling, we do suggest a very 
natural candidate.

The rest of the paper is organized as follows.  We begin with a 
general discussion of classical bulk symmetries in a field theory 
with boundaries.  Next, we apply this general discussion to the 
NLSM.  We introduce the open string NLSM by modifying the familiar 
bulk action by boundary terms,  and we find the general boundary 
conditions required for ``locality'' of this improved action.  
Next, we find the conditions for SUSY invariance of the boundary 
conditions and the action.  After giving the geometric interpretation 
of the SUSY conditions, we wrap up with a discussion of our results. 
\section{Symmetries and Boundaries in Classical Field Theory}
\subsection{Review of the classical Noether theorem.}
In order to set notation, we briefly review the connection between
symmetries and conserved charges in classical field theory.  
Consider a field theory defined on $\Sigma = \mathbb{R}^2$ by the 
action
\begin{equation}
S_\Sigma = \int_\Sigma d^2x ~ \Lc \left(\varphi^i,\d_\mu\varphi^i\right).
\end{equation}
Let $\delta_\eta\varphi^i = \eta f^i(\varphi,\d_\mu\varphi,\ldots)$ be 
an internal infinitesimal symmetry of $S_\Sigma$.  In other words, 
the variation of the Lagrangian is a boundary term:  
$\delta_\eta\Lc = -\eta \d_\mu F^\mu$ for some $F^\mu(\varphi,\ldots)$.
But, 
\begin{equation}
\delta_\eta \Lc = \eta f^i  \frac{\d \Lc}{\d \varphi^i} + \eta \d_\mu f^i
\frac{\d \Lc}{\d(\d_\mu\varphi^i)}.
\end{equation}
Hence, the current
\begin{equation}
\label{BULKJ}
j^\mu = F^\mu + f^i \frac{\d\Lc}{\d(\d_\mu\varphi^i)}
\end{equation}
is conserved up to the equations of motion.  The corresponding 
conserved charge is $Q = \int dx^1 j^0$.  That is one direction of 
the Noether theorem:  each continuous symmetry of the Lagrangian 
corresponds to the existence of a conserved current.  Now let us see
how this statement is modified in the presence of a boundary.
\subsection{Field theory on a strip}
\label{CLASSMECH}
We consider a theory with the same bulk Lagrangian but on
a strip, with $-\infty < x^0 < \infty,$ and $0 \le x^1 \le \pi.$
The addition of a boundary allows the introduction of a 
a boundary action:
\begin{equation}
S_{\d\Sigma} = \int_{\d\Sigma} dx^0 \Mc
\left(\varphi^i,\d_0\varphi^i, \d_1\varphi^i\right).
\end{equation}
In order to have a well-posed initial value problem, we must 
specify a set of boundary conditions for the fields: 
$B_n(\left.\varphi^i\right|_{\d\Sigma},\left.\d_\mu\varphi^i\right|_{\d\Sigma})
= 0$.   
The choice of $B_n$ will, in general, restrict both the values and the
variations of $\varphi^i$ on 
the boundary: the equations of motion follow by performing only
variations satisfying  $\delta B_n = 0$.\footnote{In other words, the action 
is varied over field configurations that obey the boundary conditions.} We will refer to 
this restricted set of variations as {\em allowed}
variations.  The boundary conditions are required to satisfy a 
locality constraint.  For any allowed variation, we have
\begin{eqnarray}
\delta S & = & \int_{\Sigma} d^2 x
\left[\frac{\d\Lc}{\d\varphi^i} 
              - \d_\mu\left(\frac{\d\Lc}{\d(\d_\mu\varphi^i)}\right)
\right]\delta\varphi^i\nonumber\\
~~~~~~   & ~ &+\int_{\d\Sigma} dx^0 \left\{
            \left[\frac{\d\Mc}{\d\varphi^i} 
              - \d_0\left(\frac{\d\Mc}{\d(\d_0\varphi^i)}\right) 
              + \frac{\d\Lc}{\d(\d_1\varphi^i)}\right] \delta\varphi^i
              + \frac{\d\Mc}{\d(\d_1 \varphi^i)}\d_1\delta\varphi^i\right\}.
\end{eqnarray}
If the bulk equations of motion (the first line) are to extend
smoothly to the boundary, the $B_n$ must be chosen so that the 
boundary term in $\delta S$ (the second line) is zero.  If this does not
hold, the classical equations of motion and their solutions will 
be discontinuous.  The last term, involving $\d_1\delta\varphi$, cannot
lead to local equations of motion, so we require that it vanish;
rather than imposing boundary conditions on $\d_1\varphi$ we will
restrict attention to boundary actions of the form
$\Mc(\varphi,\d_0\varphi)$.  The second line will now vanish provided
$B_n$ are chosen so that 
\begin{equation}
\label{LOCALITY}
\delta\varphi^i\left[\frac{\d\Mc}{\d\varphi^i} 
 - \d_0\left(\frac{\d\Mc}{\d(\d_0\varphi^i)}\right) 
 + \frac{\d\Lc}{\d(\d_1\varphi^i)}\right] = 0 
\end{equation}
for any allowed variation.

One may be tempted to think of $B_n$ as 
``boundary equations of motion.''  This is quite misleading, as the 
boundary conditions are much stronger than equations of motion.  In 
quantizing the problem via path integral techniques, we expect that 
the boundary conditions are to be imposed on all field configurations.
Thus, the boundary conditions hold as operator equations, and  they
can be used in the boundary action.  

The symmetry of the bulk theory will persist in the presence of
boundaries if (in classical mechanics) the symmetry transform of a
classical trajectory is another classical trajectory.  To define the
symmetry transform, the variation $\delta_\eta\varphi^i = \eta f^i$ must be an
allowed variation about any classical trajectory.  Boundary conditions
for which a given symmetry variation is allowed in this sense will be
called {\em classically symmetric} boundary conditions.  Classical
trajectories will transform into others if the equations of motion are
invariant under the symmetry, which means the symmetry variation of the 
action vanishes.

The symmetry variation of the action is given by 
\begin{equation}
\delta_\eta \left(S_\Sigma + S_{\d\Sigma}\right) = \intb \left( -\eta F^1 +
\eta f^i\left\{\frac{\d\Mc}{\d\varphi^i} -
\d_0\left(\frac{\d\Mc}{\d(\d_0\varphi^i)}\right)\right\} \right).
\end{equation}
Suppose that the boundary conditions are symmetric so that the symmetry
variation is an allowed variation. Then we can use Eqs. (\ref{BULKJ},\ref{LOCALITY}), 
to conclude that 
\begin{equation}
\delta_\eta S = - \eta \intb ~j^1.
\end{equation}
If the action is invariant, then 
$\left. j^1\right|_{\d\Sigma} = \d_0 K,$ for some 
$K(\varphi,\d_0\varphi)$  defined on the boundary.
It is then clear that $\Qt = Q+K$ is conserved:
\begin{equation}
\d_0 \Qt = \d_0 \int dx^1 j^0 + \left.\d_0 K\right|_{\d\Sigma} = -\left.
j^1\right|_{\d\Sigma} + \left.\d_0 K\right|_{\d\Sigma} = 0.
\end{equation}

Let us contrast  $K$ with a similar bulk quantity, $F^\mu$. 
The crucial difference between these two is that the latter is determined 
by the bulk action, while the former is determined by the choice of boundary 
conditions.  The presence of a boundary will, in general, break the bulk 
symmetry.  By choosing boundary conditions appropriately, one can ensure that 
the symmetry is preserved, with the conserved charge given as above: 
$\Qt = Q + K$.  In principle, one could imagine that different boundary 
conditions could lead to the same preserved bulk symmetry but with a 
different conserved charge: $\Qt' = Q + K'$.

With this background in mind, we will now proceed to the $\Nc =
(2,2)$ NLSM.  We will add boundaries, consider locality, and ensure 
that the boundary conditions are supersymmetric under an 
$\Nc = 2$ subalgebra.
\section{The NLSM with Boundaries}
 \subsection{The Action}
The NLSM is a field theory of maps $\Phi: \Sigma \rightarrow X$ from
the worldsheet $\Sigma$ to the target space $X$.  Locally (both on the
worldsheet and target space), we can specify such a map by a set of 
functions $\phi^I(x): \Sigma \rightarrow \mathbb{R} $, $I = 1 \ldots d$, 
where $d$ is the (real) target space dimension, and we can think of 
$\phi^I$ as coordinates on a patch of the target space.  In order for 
the action to have $\Nc = (1,1)$ supersymmetry, we must add 
worldsheet Majorana-Weyl fermions, $\psi_\pm^I(x)$, sections of
$K^{\pm \frac{1}{2}}\otimes \Phi^*(TX)$, where $K$ is the canonical
bundle of $\Sigma$,  and $\Phi^*$ is the pull-back by the map $\Phi:
\Sigma \rightarrow X$. We take our action to be $S = \Sbu + \Sbo.$
The bulk piece is the familiar closed string $\Nc = (2,2)$ NLSM: 
\begin{eqnarray}
\label{SBU}
\Sbu & = & \int_\Sigma d^2 x \Bigl\{ \frac{1}{2} G_{IJ} \bigl[ -\d_\mu
\phi^I \d^\mu \phi^J 
                           +i\left( \psi_-^I D_+ \psi_-^J + \psi_+^I D_-
\psi_+^J \right) \bigl] \nonumber\\
 ~~~ & ~~ & ~~~~~~~~~ + \frac{1}{4} R_{IJKL}~\psi_+^I \psi_+^J \psi_-^K
\psi_-^L 
                      - \frac{1}{2} B_{IJ} \ep^{\mu\nu} \d_\mu \phi^I
\d_\nu \phi^J \Bigr\},
\end{eqnarray}
We work on a worldsheet with a flat Minkowski metric of signature $(-,+)$
and $\ep^{01} = +1$ .  We define $\d_\pm = \d_0 \pm \d_1.$  The covariant 
derivatives are defined by 
\[
D_\pm \psi^I = \d_\pm \psi^I + \d_\pm \phi^J \Gamma^I_{JK} \psi^K.
\] 
The $\Gamma^I_{JK}$ are Christoffel symbols for the Levi-Civita
connection associated with the target space metric.
If the target space metric is K\"ahler and $H = d B = 0$, this action
has $\Nc = (2,2)$ supersymmetry. In addition, this NLSM is classically 
conformally invariant.  One should keep in mind that in
the full string theory, the model would include free fields
representing the noncompact directions in spacetime.  Since the
worldsheet field theory factorizes, we will restrict attention to the
internal degrees of freedom.

Now we consider the boundary action $\Sbo$.  The most general boundary
action invariant (classically) under boundary reparametrizations and 
scale transformations is 
\begin{eqnarray}
\label{SBO}
\Sbo  = \int_{\d \Sigma} dx^\mu V_\mu(\phi,\psi_-,\psi_+) + \int_{\d
\Sigma} \sqrt{\left|dx^\mu dx^\nu \eta_{\mu\nu}\right|}~W
(\phi,\psi_+,\psi_-),
\end{eqnarray}
where
\begin{eqnarray}
\label{VPM}
V_- & = & A_I(\phi) \d_- \phi^I + i C_{IJ}(\phi) \psi_-^I \psi_-^J,
\nonumber\\
V_+ & = & A_I(\phi) \d_+ \phi^I + i \Ct_{IJ}(\phi) \psi_+^I \psi_+^J,
\nonumber\\
W   & = & i D_{IJ}(\phi) \psi_+^I\psi_-^J.
\end{eqnarray}
The $A_I(\phi)$, $C_{IJ}(\phi)$, $\Ct_{IJ}(\phi), D_{IJ}$ are tensors on
the target space.\footnote{The same field $A_I(\phi)$ couples to $\d_+\phi$ 
and $\d_-\phi$ to ensure that only the tangential derivative $\d_0 \phi$
appears in the boundary action.} If we take the worldsheet to be a strip 
as above, then
\begin{equation}
\Sbo = \left. \int_{-\infty}^{\infty} dx^0 \left\{ W + \frac{1}{2} (V_+
+ V_-) \right\}\right|_{x^1 = 0}^{x^1 = \pi}.
\end{equation}
We will restrict attention to this case in what follows.
 \subsection{The Boundary Conditions}
Let us vary the action with respect to $\phi$ and $\psi_\pm$. 
In computing the variation it is important to note that when we vary
the map $\Phi$ the Fermi fields, which as noted above are sections of
bundles determined by $\Phi$, cannot be held ``constant.''  We can
think of $\psi_\pm$ as sections of a bundle over the space of maps
$\Sigma\to X$ with connection given by pulling back $\Gamma$. Parallel
transport then determines the variation
\begin{equation}
\label{PSIVAR}
\delta \psi_\pm^I =   \dpsit^I_\pm + \Gamma^I_{~JK}\psi_\pm^J\delta
\phi^K,
\end{equation}
where $\dpsit_\pm$ is the variation of $\psi_\pm$ independent of
$\delta\phi$.
\begin{eqnarray}
\label{SVAR1}
\delta S & = &~~~\int_\Sigma d^2x \left\{ \delta \phi^K E_K + \dpsit_+^I
E^+_I +\dpsit_-^I E^-_I \right\}\nonumber \\
 ~       & ~ & + \intb \left\{ F_{KI}\d_0 \phi^I - G_{KI} \d_1 \phi^I
+\frac{1}{2} V'_K \right\} \delta \phi^K   \nonumber\\
~        & ~ & + \frac{i}{2}\intb \Bigl\{ \dpsit_-^I\left[\left(2 C_{IJ}
- G_{IJ}\right)\psi_-^J - 2 D_{JI}\psi_+^J\right]  \Bigr. \nonumber\\
~        & ~ & ~~~~~~~~~~~~~~ \Bigl. +\dpsit_+^I\left[\left(2\Ct_{IJ}+
G_{IJ}\right)\psi_+^J +2 D_{IJ} \psi_-^J \right] \Bigr\}
\end{eqnarray}
where  $F_{KI} = A_{I,K} - A_{K,I} + B_{KI}$, 
$V'_K = i\left(   C_{IJ;K}\psi_-^I\psi_-^J +2 D_{IJ;K}\psi_+^I\psi_-^J 
+\Ct_{IJ;K}\psi_+^I\psi_+^J\right)$,
and $C_{IJ;K} = \nabla_K C_{IJ}$.
The bulk term corresponds to the bulk equations of motion for the
fields:
\begin{eqnarray}
\label{EOM}
E^I & = & D^2 \phi^I -\frac{i}{2} R^I_{~JKL}\left(\d_-\phi^J\psi_+^K\psi_+^L
                                    + \d_+\phi^J\psi_-^K\psi_-^L\right) 
                -\frac{1}{4} G^{IA}
R_{JKLM;A}\psi_+^J\psi_+^K\psi_-^L\psi_-^M, \nonumber\\
E^-_I & = & D_+\psi_{-I} - \frac{i}{2} R_{IJKL} \psi_-^J\psi_+^K\psi_+^L,\nonumber\\
E^+_I & = & D_-\psi_{+I} - \frac{i}{2} R_{IJKL} \psi_+^J\psi_-^K\psi_-^L.
\end{eqnarray}
As discussed in the previous section, we must choose boundary conditions
such that the boundary term in $\delta S$ vanishes.  We use the standard 
Ansatz for the fermion boundary conditions:
\begin{equation}
\label{FERMB}
\psi_+^I = \Rt^I_{~J}(\phi) \psi_-^J.
\end{equation}
As previously noted in \cite{LZ}, this form is
unique provided that we demand that it respects classical conformal 
invariance and is non-singular in field-space.  These boundary conditions
constrain the variations of $\psi_\pm.$  The allowed variations must
satisfy
\begin{equation}
\label{FERMBVAR}
\dpsit_+^I = \Rt^I_{~J;K}\delta\phi^K \psi_-^J + \Rt^I_{~J} \dpsit_-^J.
\end{equation}
Plugging these expressions into the variation, we find that the boundary
term takes the form
\begin{eqnarray}
\label{SVAR2}
\delta S  & = & \intb \left\{ \left[ F_{KI}\d_0 \phi^I - G_{KI} \d_1
\phi^I +\frac{1}{2} V_K \right] \delta \phi^K 
                + \frac{i}{2}\dpsit_-^I\Bigl[2 C_{IJ} - G_{IJ} 
\Bigr.\right.\nonumber\\
~         & ~ & ~~~~~~~\left. \left.+ \Rt^K_{~I}\left(2\Ct_{KL} +
G_{KL}\right)\Rt^L_{~J}
                             +2 \Rt^K_{~I} D_{KJ} -2 \Rt^K_{~J}
D_{KI}\right]\psi_-^J \right\},
\end{eqnarray}
where 
\begin{eqnarray}
V_K & = & i \left\{ C_{IJ;K} + R^L_{~J;K}\left(2 D_{LJ} + \left(2 \Ct_{LM} +
G_{LM}\right) R^M_{~J}\right) \right. \nonumber \\
    & ~ & \left. + 2 R^L_{~I} D_{LJ;K} + R^L_{~I} C_{LM;K} R^M_{~J}\right\}\psi_-^I\psi_-^J 
\end{eqnarray}
Since we do not wish to constrain $\psi_-$ on the boundary, 
we find that in order for locality to hold we must have
\begin{eqnarray}
\label{FERMLOC}
\Rt^T G \Rt & = & G, \nonumber\\
\Ct & = & \Rt D^T - D \Rt^T - \Rt^T C \Rt.
\end{eqnarray}
We use an obvious notation: $\Rt^A_{~I} G_{AB} \Rt^B_{~J} = G_{IJ}$ is written 
as $\Rt^T G \Rt = G$, etc.
We now need to choose boundary conditions for the bosons.  We assume that the 
end-point of the string moves along $M$, a submanifold of $X$.  This represents
a $D$-brane wrapping $M \subseteq X.$ Thus, the allowed $\delta\phi$
are tangent to $M$ on the boundary, and the boson boundary conditions
we need to impose for locality are
\begin{equation}
\label{BOSLOC1}
\d_1 \phi^K = F^K_{~I}\d_0 \phi^I + \frac{1}{2} V^K - n^K,
\end{equation}
everywhere on $\d\Sigma$, where $n^K$ is an arbitrary vector
normal to $M$.  In an open neighborhood $U \subset M$ of $\Phi(z)$ for
any $z\in\d\Sigma$ we can choose a basis $e^{~I}_{\alpha}$ for
$\left.TX\right|_U$ (a vielbein) adapted to the 
splitting of $\left.TX\right|_{U}=TM\oplus NM$ into tangent and normal
directions.  We split the index $\alpha$ into tangential indices 
(labelled by $a,b,c,\ldots$) and normal indices (labelled by 
$z,y,x,\ldots$).  In this basis the boson boundary conditions are
\begin{equation}
\label{BOSB}
\d_+\phi^\alpha = R^\alpha_{~\beta} \d_-\phi^\beta + T^\alpha_{~\beta}
V^\beta,
\end{equation}
where $R$ is an orthogonal matrix 
\begin{equation}
\label{RMATRIX}
R = \left( \begin{array}{cc}      
                   \left(\frac{1+F}{1-F}\right)_{ab} &   0 \\
                             0          & -\delta_{zy} 
           \end{array}
     \right),
\end{equation}
and $T_{\alpha\beta} = \frac{1}{2} \left(\delta_{\alpha\beta} +
R_{\alpha\beta} \right).$ It
will be important later that in the vielbein, the only nonvanishing 
components of $T$ are $T_{ab} = (1-F)^{-1}_{ab}$.  Furthermore, the 
daunting $V_\alpha$ is greatly simplified by the use of the fermion 
boundary conditions and Eq.  (\ref{FERMLOC}):
\begin{equation}
\label{VFORM}
V_{\alpha} = i \Rt_{\mu\beta;\alpha} \Rt_{\mu\delta}
\psi_-^\beta\psi_-^\delta.
\end{equation}
Note that $V$ is independent of the boundary fermion couplings
$C$,$\Ct$, and $D.$ This reflects the general result that once the 
fermion boundary conditions are imposed, then the fermion bilinear 
term in the boundary action is identically zero, hence trivially SUSY
invariant, for any $C$, $\Ct,$ $D$ satisfying Eq.(\ref{FERMLOC}).  
The locality constraint Eq.(\ref{FERMLOC}), which can be solved for
$\Ct$ in terms of $C$, $D$, and $\Rt$, is thus the only constraint 
on these couplings.
As we will see shortly, $\Rt$ is fixed by supersymmetry.
\section{Conditions for $\Nc = 2$ Supersymmetry}
In this section we find the restrictions on $M$ and the boundary couplings
which lead to an unbroken $\Nc = 2$ SUSY.  We will express these as
equations to be satisfied everywhere along $M$ by the various
background fields.  In the next section we will study these equations
and interpret them geometrically.

\subsection{SUSY Variation of the NLSM Fields}
The bulk action is supersymmetric up to a boundary term under $\Nc
= (2,2)$ supersymmetry.  Since the target space is a complex manifold, 
and we are working locally, we can choose a set of coordinates where the 
complex structure $J^A_{~B}$ is constant, in addition to being covariantly 
constant.  In these coordinates the SUSY variations take a particularly 
simple form:
\begin{eqnarray}
\label{SUSYVAR0}
\delta \phi^I  & = & i \left(\epp^2 \psi_-^I + \epp^1 J^I_{~J} \psi_-^J
- \epm^2 \psi_+^I -\epm^1 J^I_{~J} \psi_+^J\right),\nonumber \\
\delta\psi_-^I & = & \dpsit_-^I  +\Gamma^I_{~JK} \psi_-^J\delta\phi^K,
\nonumber\\
\delta\psi_+^I & = & \dpsit_+^I  +\Gamma^I_{~JK} \psi_+^J\delta\phi^K, 
\end{eqnarray}
where
\begin{eqnarray}
\label{SUSYVAR01}
\dpsit_-^I & = & -\epp^2\d_-\phi^I + \epp^1 J^I_{~K} \d_-\phi^K,
\nonumber\\
\dpsit_+^I & = & \epm^2\d_+\phi^I - \epm^1 J^I_{~K} \d_+\phi^K. 
\end{eqnarray}

Up to an irrelevant phase, there are two choices for the $\Nc = 2$
subalgebra of $\Nc =(2,2)$ SUSY preserved by the boundary.\footnote{
Essentially, these are the two sets of SUSY generators that 
anticommute to $\d_0,$ the unbroken translation generator.}  These are 
commonly labelled as {\bf A} and {\bf B}, and they are parametrized by $\ep$ 
and $\ep'$, with the $\Nc = (2,2)$ parameters given by 
\begin{eqnarray}
\ep_+^2 & = & -\ep_-^2 = \ep, \nonumber\\
\ep_+^1 & = & \eta \ep_-^1 = \ep',
\end{eqnarray}
where $\eta = +1 $ for {\bf A} SUSY and $\eta = -1$ for {\bf B} SUSY.

The {\bf A}/{\bf B} SUSY variations are 
\begin{eqnarray}
\label{SUSYVAR1}
\delta\phi^I & = &   i\ep         \left(\psi_-^I +     \psi_+^I\right) 
              + i\ep' J^I_{~J}\left(\psi_-^J -\eta
\psi_+^J\right),\nonumber\\
\dpsit_-^I & = & -\ep \d_-\phi^I + \ep' J^I_{~K}\d_-\phi^K,\nonumber\\ 
\dpsit_+^I & = & -\ep \d_+\phi^I -\eta \ep' J^I_{~K}\d_+\phi^K
\end{eqnarray}
 \subsection{SUSY of the Fermion Boundary Condition}
Now we will study the supersymmetry variations of the fermion
boundary conditions.  Equating coefficients of expressions in the
independent boundary fields, invariance of the boundary 
conditions will lead to geometric constraints on background fields. 
\subsubsection{The $\ep$ variation}
Plugging the $\ep$ SUSY variation into Eq. (\ref{FERMBVAR})
we get the condition
\begin{equation}
-(\d_+\phi^I - \Rt^I_{~J}\d_-\phi^J) = 2 i \Rt^I_{~J;K} T^K_{~L}
\psi_-^L\psi_-^J.
\end{equation}
Using the boson boundary conditions, Eq. (\ref{BOSB}), we find
\begin{equation}
-\left(R^I_{~J} - \Rt^I_{~J}\right)\d_-\phi^J = T^I_{~J} V^J - 2 i
\Rt^I_{~J;K} T^K_{~L} \psi_-^J\psi_-^L.
\end{equation}
Since we do not wish to impose any further constraints on $\d_-\phi^I$
and $\psi_\_^I,$ the coefficients of $\d_-\phi^I$ and $\psi_-^J\psi_-^L$ 
must vanish separately. 
Thus,
\begin{equation}
\label{FixRt}
\Rt = R.
\end{equation}
The rest of the computation is simplified if we replace $\Rt$ by 
$R$, so we will do so.

The vanishing of the $\psi_-\psi_-$ term implies
\begin{equation}
\label{FERMCOND1}
-T_{\alpha a} V_a = 4 i T_{\alpha\beta; a} T_{a\delta} 
                       \psi_-^\delta\psi_-^\beta,
\end{equation}
where we have used $R^I_{~J;K} = 2 T^I_{~J;K},$ and written 
the expression in the vielbein basis. 

To avoid confusion, let us be explicit about the form of the covariant
derivatives in the vielbein basis:
\[ A_{\alpha;\beta} = A_{I;J} e^I_{\alpha} e^J_{\beta}. \]  More
explicitly,
\[ A_{\alpha;\beta} = A_{\alpha,\beta} + \omega_{\beta\alpha\gamma}
A_\gamma,\]
where $A_{\alpha,\beta} = e^I_\beta \nabla_I A_{\alpha},$ and 
$\omega_{\beta\alpha\gamma} = e^I_{\beta} e^J_{\alpha} \nabla_I
e_{\gamma J}$ is the spin connection.

Using the form of $V_\alpha$ given in Eq. (\ref{VFORM}), we find the
following
restriction on $R_{\alpha\beta}:$
\begin{equation}
\label{FERMCOND2}
T_{\alpha a} \left(T_{\mu\delta;a} R_{\mu\beta} - T_{\mu\beta;a}
R_{\mu\delta}\right) = 
                        2 \left(T_{\alpha\delta;a} T_{a\beta} -
T_{\alpha\beta;a} T_{a\delta}\right).
\end{equation}
To extract the consequences of this equation, we split the free
indices into the normal and tangent directions, and carry out 
a case-by-case analysis.  This is aided by a few helpful and easily 
verifiable facts:
\begin{eqnarray}
\label{TPROPERTIES}
T_{xy;a} & = & 0 ,\nonumber\\
T_{xb;a} & = & \omega_{axc} T_{cb}, \nonumber \\
T_{bx;a} & = & \omega_{axc} T_{bc}, \nonumber \\
T_{ab;c} & = & T_{ad} H_{dec} T_{eb},
\end{eqnarray}
where 
$H_{dec} = -H_{edc} = F_{de;c} + \omega_{czd} F_{ze} - \omega_{cze}
F_{zd}$
Also, the number of cases to consider is lessened if we note the
antisymmetry of Eq. (\ref{FERMCOND2}) in the $\beta,\delta$ indices.  
We find that Eq. (\ref{FERMCOND2}) is equivalent to
\begin{equation}
 \label{FERMSUSYCOND1}
 \omega_{eza} = \omega_{aze},
\end{equation}

\begin{equation}
 \label{FERMSUSYCOND2}
 T_{ae}\left(T_{fd;e}R_{fb} - T_{fb;e} R_{fd}\right)
 = 2\left( T_{ad;e} T_{eb} - T_{ab;e} T_{ed}\right).
\end{equation}

So, the fermion boundary conditions are invariant under the $\ep$ SUSY if
and only if these two requirements are met. 
\subsubsection{The $\ep'$ variation }
Plugging the $\ep'$ variation into Eq.  (\ref{FERMBVAR}) and 
using the boson boundary conditions, we find
\begin{equation}
 \label{FERM2COND1}
-\left(\eta J^I_{~J} R^J_{~K} + R^I_{~J} J^J_{~K}\right)\d_-\phi^K =
 \eta J^I_{~K} T^K_{~J} V^J +4i T^I_{~J;K} T^K_{~L} J^L_{~M} \psi_-^M\psi_-^J.
\end{equation}
Again, the two sides must vanish separately.  The left-hand side
gives the condition
\begin{equation}
 \label{RJ}
  R J + \eta J R = 0.
\end{equation}
The left-hand side gives
\begin{equation}
\eta J_{\alpha e} T_{ef} T_{\mu\delta;f} R_{\mu\beta} 
- 2 T_{\alpha\delta;e} T_{ef} J_{f\beta} = \left( \delta \leftrightarrow
\beta \right).
\end{equation}
It turns out that it is not necessary to do the case-by-case analysis as
above.  Using
$J^2 = -1$, $JR +\eta RJ = 0$, and the covariant constancy of $J$, 
it is easy to show
that this is equivalent to Eq. (\ref{FERMSUSYCOND2}).  Thus,
$\Nc = 2$ invariance follows from $\Nc = 1$ and Eq. (\ref{RJ}).
 \subsection{SUSY of the Boson Boundary Condition}
Now we explore the supersymmetry of the boson boundary conditions.  As a
warm-up, let us consider 
what happens in flat space with $F=0$. This is an easy case, since the
matrix $R$ is now constant,
and the two fermion term in the boson boundary conditions can be taken to
be zero.
Consider the $\ep$ SUSY, for which $\delta \phi^I = i\ep\left(\psi_-^I+
\psi_+^I\right).$  
Plugging this variation into the bosonic boundary conditions, we have the
requirement
\begin{equation}
\d_+\left(\psi_-^I+\psi_+^I\right) = R^I_{~J} \d_-\left(\psi_-^J +
\psi_+^J\right).
\end{equation}
This is not an algebraic condition on the $\psi_\pm$, and we cannot
satisfy it by using just the
algebraic conditions we already have.  One is tempted to differentiate
the fermion boundary
conditions.  Since $\psi_+^I(x^0,0) = R^I_{~J}
\psi_-^J(x^0,0)$ must hold for all $x^0$, we are
allowed to differentiate this relation with respect to $x^0$. 
Unfortunately, it does not make sense to differentiate 
it with respect to $x^1$. We recall, however, the fermion equations of
motion:  $\d_\mp \psi_\pm^I = 0,$
which allow us to relate $\d_1 \psi_\pm$ to $\d_0\psi_\pm$.  By using
the equations of motion, we see
that the variation of the bosonic boundary conditions becomes
\begin{equation}
\d_0 \psi_+^I = R^I_{~J} \d_0 \psi_-^J,
\end{equation}
which is satisfied if the fermion boundary conditions are
supersymmetric.  Classically, the use of the equations of motion
is justified.  Since we are eventually interested in the quantum
problem, we will need to find a suitable modification.  We discuss
this below.  Here we will show that the boson boundary conditions are 
supersymmetric up to the fermion equations of motion. 

The computation is greatly simplified if we use the identity 
$T^I_{~J} V^J = 4i T^I_{~J;K} T^K_{~L} \psi_-^J \psi_-^L$.  This 
reduces to Eq.(\ref{FERMCOND1}) when we use the properties of $T$.
Since this holds for any values of the Fermi fields satisfying the
(SUSY) boundary conditions, its SUSY variation is also an identity.
We will use this to write the boson boundary conditions as
\begin{equation}
\label{BOSBCMOD}
\d_+\phi^I - R^I_{~J} \d_-\phi^J = i S^I_{~JL} \psi_-^J \psi_-^L,
\end{equation}
where $S^I_{~JL} = 2\left(T^I_{~J;K} T^K_{~L} - T^I_{~L;K}
T^K_{~J}\right).$
The variation will take the form
\begin{equation}
\label{BOSBCMODVAR}
\d_+\delta\phi^I - R^I_{~J,K}\delta\phi^K\d_-\phi^J -
R^I_{~J}\d_-\delta\phi^J = 
i \delta\left(S^I_{~JL}\psi_-^J\psi_-^L\right).
\end{equation}
\subsubsection{The $\ep$ variation}
Let us first work out the left-hand side of Eq. (\ref{BOSBCMODVAR}).
\begin{equation}
\label{LHS1}
\LHS = i\ep\left\{\d_+ \left(\psi_-^I + \psi_+^I\right) -
R^I_{~J}\d_-\left(\psi_-^J+\psi_+^J\right)\right\}
       -2i\ep R^I_{~J,K}\d_-\phi^J T^K_{~L} \psi_-^L.
\end{equation}
We will use the fermion equations of motion (Eq.(\ref{EOM})) to simplify
the $\left\{\ldots\right\}$ above. 
We find
\begin{eqnarray}
\label{LHSep2}
\LHS & = & i\ep\left\{2 S^I_{~JL}\psi_-^J\d_-\phi^L \right.\nonumber\\
~   & ~ & + i \left[ \left(R^I_{~C;L} - 2 \Gamma^I_{LJ}
T^J_{~C}\right)S^L_{AB}\right.\nonumber\\
~   & ~ & +T^I_{~J} R^J_{~DEF}\left.\left(\delta^D_{~A}R^E_{~B} R^F_{~C}
- R^D_{~A} \delta^E_{~B} \delta^F_{~C}\right)\right]
          \psi_-^A\psi_-^B\psi_-^C\left.\right\}.
\end{eqnarray}
The variation of the right-hand side is straightforward.
\begin{equation}
\label{RHSep2}
\RHS = 2i\ep \left\{S^I_{~JL}\psi_-^J\d_-\phi^L + i \left(S^I_{~AB;K} -
\Gamma^I_{~KJ} S^J_{~AB}\right)T^K_{~C}\psi_-^A\psi_-^B\psi_-^C\right\}.
\end{equation}
Setting $\LHS = \RHS$, we see that the the one fermion terms cancel, and
we are left with the three fermion terms.  
The boson boundary conditions are invariant under the $\ep$ SUSY variation
if 
\begin{equation}
\label{BOSONCOND1}
Q^I_{~[ABC]} = \frac{1}{2} T^I_{~J}
R^J_{~DEF}\left(\delta^D_{~[A}R^E_{~B} R^F_{~C]} -
R^D_{~[A}\delta^E_{~B}\delta^F_{~C]}\right),
\end{equation}
where $Q^I_{~ABC} = S^I_{~AB;K} T^K_{~C} - S^K_{~AB} T^I_{~C;K}.$  
We will spare the reader the details, but roughly, the equality can
be shown as follows. There are terms of the form $[\nabla_L,\nabla_M] T$ 
in $Q^I_{~[ABC]}.$ By the definition of the Riemann tensor, these can be 
written as sums of contractions of the  Riemann tensor with $T$.  Then
Eq. (\ref{BOSONCOND1}) can be shown to hold by repeated use of 
$R_{I[JKL]} = 0$.  We see that, up to the equations of motion,  there are 
no new constraints from the SUSY of the boson boundary conditions.
\subsubsection{The $\ep'$ variation }
We begin the same way as for the $\ep$ variation.  
 Eliminating the $\d_1\psi_\pm$ by the equations of motion, and using the
boundary conditions, we find that the variation of the left-hand
side is
\begin{eqnarray}
\label{LHS2}
\LHS & = & 2i\ep'\left\{S^I_{~AD} J^A_{~B} \d_-\phi^D \psi_-^B 
                     + i\left[\left(T^I_{~J;L} -
\Gamma^I_{~LK}T^K_{~J}\right) J^J_{~A}
S^L_{~BC}\right.\right.\nonumber\\
~    & ~ & +\left.\left.\frac{1}{2} T^I_{~J} J^J_{~K} R^K_{~DEF}
            \left(\delta^D_{~A}R^E_{~B}R^F_{~C}+\eta
R^D_{~A}\delta^E_{~B}\delta^F_{~C}\right)\right]\right\}.
\end{eqnarray}
The variation of the right-hand side gives
\begin{eqnarray}
\label{RHS2}
\RHS = & = & 2i\ep'\left\{S^I_{~AB}J^A_{~D} \d_-\phi^D\psi_-^B + 
                           i\left(S^I_{~AB;K} - \Gamma^I_{~KL}
S^L_{~AB}\right)T^K_{~M}J^M_{~C}\psi_-^A\psi_-^B\psi_-^C\right\}
\end{eqnarray}

To show that $\LHS = \RHS$, we will first show that the one fermion
terms match.  This is so if 
\begin{equation}
\label{JSJ}
S_{\gamma\alpha\beta} = J_{\rho\alpha} S_{\gamma\rho\nu} J_{\nu\beta}.
\end{equation}
But, SUSY of the fermion boundary conditions implies
(Eq.(\ref{FERMCOND2}))
\begin{equation}
S_{\gamma\rho\nu} = T_{\gamma e} \left(T_{\mu\rho;e} R_{\mu\nu} -
T_{\mu\nu;e} R_{\mu\rho}\right).
\end{equation}
Using $T_{\mu\lambda;e} = \eta J_{\mu\chi} T_{\chi\kappa;e}
J_{\kappa\lambda},$ and $JR+\eta RJ=0$, we find
\begin{equation}
J_{\rho\alpha}S_{\gamma\rho\nu} J_{\nu\beta}  = T_{\gamma e}
\left(T_{\mu\alpha;e} R_{\mu\beta } - T_{\mu\beta;e}
R_{\mu\alpha}\right) = S_{\gamma\alpha\beta}
\end{equation}
So, we see that the one fermion terms do indeed agree.

What about the three fermion terms?  The remaining condition to be
satisfied is
\begin{eqnarray}
\left(S^I_{~AB;K}T^K_{~J} -T^I_{~J;L} S^L_{~AB}\right)\psi_-^A\psi_-^B
J^J_{~C}\psi_-^C = \nonumber\\
\frac{1}{2}
T^I_{~J}J^J_{~K}R^K_{~DEF}\left(\delta^D_{~A}R^E_{~B}R^F_{~C}+\eta
R^D_{~A}\delta^E_{~B}\delta^F_{~C}\right)
                               \psi_-^A\psi_-^B\psi_-^C.
\end{eqnarray}
Using $S= JSJ$, the left-hand side takes the form
\begin{equation}
\LHS = \left(S^I_{~AB;K} T^K_{~C} - T^I_{~C;K}
S^K_{~AB}\right)\Ph^A\Ph^B\Ph^C,
\end{equation}
where $\Ph^A = J^A_{~J} \psi_-^J.$

With a little work we can rewrite the right-hand side as
\begin{equation}
\RHS = \frac{1}{2} R^I_{~DEF}\left(\delta^D_{~A}R^E_{~B} R^F_{~C} -
R^D_{~A}\delta^E_{~B}\delta^F_{~C}\right)\Ph^A\Ph^B\Ph^C.
\end{equation}
Setting $\LHS = \RHS$, and extracting the piece totally antisymmetric in
$(ABC),$ we find Eq.(\ref{BOSONCOND1}).  

So, we have shown that, up to the fermion equations of motion, the boson
boundary condition is SUSY
provided that the fermion boundary condition is SUSY.
 \subsection{SUSY of the Action}
We will now examine the conditions for the action to be supersymmetric.
This computation is simplified by the realization that the fermion
bilinear boundary couplings drop out of the action once we use the 
boundary conditions.

Thus, 
\begin{equation}
\label{SBOUNVAR}
\delta S_{\d\Sigma} =  -\frac{i}{2} \intb ~ 2
\left(\d_+\phi^I+\d_-\phi^I\right) F_{IK}
                                   \left[\ep T^K_{~L} +\ep' T^K_{~M}
J^M_{~L}\right]\psi_-^L
\end{equation}
The variation of the bulk action is given by:
\begin{equation}
\label{SBULKVAR}
\delta S_{\Sigma} = \frac{i}{2} \intb \left\{\ep\left(\d_-\phi^I R_{IL}
- \d_+\phi^I G_{IL}\right) -
                                             \ep'\left(\eta\d_-\phi^I
J_{IJ} R^J_{~L} + \d_+\phi^I J_{IL}\right)
                                      \right\}\psi_-^L.
\end{equation}
\subsubsection{The $\ep$ variation}
Extracting the $\ep$ term from above, and using the boson boundary
conditions, we find
\begin{eqnarray}
\delta S & = & \frac{i\ep}{2}\intb \left\{\d_-\phi^I\left(-4T^J_{~I}
F_{JK} T^K_{~L} + R_{IL} - R_{LI}\right)
               \d_-\phi^I\psi_-^L \right.\nonumber\\
~        & ~ & ~~~~~~~~~~~~~~~\left. -V^I T^J_{~I}\left(G_{JL} + 2
F_{JK} T^K_{~L}\right)\right\}.
\end{eqnarray}
It is easy to show that $R - R^T = 4 T^T F T$, so that the one fermion
term is zero.  Since, $T^T(1+ 2 FT) = T,$
it follows that
\begin{equation}
\delta S = -\frac{i\ep}{2} \intb   V_I T^I_{~L} \psi_-^L.
\end{equation}
Using the expression for $V$ (Eq.(\ref{VFORM})) the condition for the
action to be invariant under
the $\ep$ SUSY is
\begin{equation}
\label{JUSTABOVE}
T_{e\alpha} T_{\mu\beta;e} R_{\mu\delta}
\psi_-^\alpha\psi_-^\beta\psi_-^\delta = 0.
\end{equation}
This leads to one non-trivial requirement:
\begin{equation}
\label{SSUSYCOND1}
T_{cf;e}\left[T_{ea}\delta_{fb} R_{cd} + (bda) + (dab)\right] = 0.
\end{equation}
\subsubsection{The $\ep'$ variation}
The $\ep'$ variation of the action is
\begin{eqnarray}
\delta S & = & \frac{i\ep'}{2}\intb \left\{\d_-\phi^I\left(-4 T^J_{~I}
F_{JK} T_{KM} + R_{IM} - R_{MI}\right) 
                                                        J^M_{~L}\psi_-^L
\right. \nonumber\\
~        & ~ & ~~~~~~~~~~~ \left.    -V^I T^J_{~I}\left(G_{MJ} + 2
J_{JK} T^K_{~M}\right)J^M_{~L}\psi_-^L\right\}.
\end{eqnarray}
As in the $\ep'$ variation, the one fermion term is zero, and we are
left with the condition
\begin{equation}
T_{a\gamma} T_{\mu\beta;a} R_{\mu\delta} J_{\gamma\alpha} \psi_-^\alpha
\psi_-^\beta \psi_-^\delta= 0.
\end{equation}
To show that this holds, we use $JR+\eta RJ = 0$, and $ (JTJ)_{;a} =
\eta T_{;a}.$  The condition becomes
\begin{equation}
T_{\mu\alpha;a} R_{\mu\beta} T_{a\gamma} \Ph^\alpha\Ph^\beta\Ph^\gamma
=0,
\end{equation}
which is satisfied provided that Eq. (\ref{JUSTABOVE}) holds.
\section{Satisfying the Constraints}
 \subsection{Algebraic Conditions on $R,J,F$}
Here we will explore the geometric meaning of the condition $RJ + \eta
JR = 0.$  
Using the explicit form of $R$ in the vielbein (Eq. (\ref{RMATRIX})),we
find that for {\bf A}-type supersymmetry, 
$J$ and $F$ must satisfy the following:
\begin{eqnarray}
\label{COI}
J_{zy}  & = & 0, \nonumber\\
J_{ab}  & = & F_{ac} J_{cd} F_{db}, \nonumber\\
F_{ab} J_{bx} & = & 0.
\end{eqnarray}
The first equation implies that the {\bf A}-type supersymmetric cycle 
is locally a co-isotropic submanifold.   The last two equations have
solutions if and only if $\dim M = \frac{1}{2} \dim X + 2 k,$ where $k$ 
is a non-negative integer.  A cycle is Lagrangian if and only if $F_{ab} = 0.$  
Note that if $B\neq 0,$ $F$ is not a curvature associated to a connection 
on a line bundle!

For {\bf B}-type supersymmetry, we find a different set of conditions:
\begin{eqnarray}
\label{HOLO}
J_{az} = 0, \nonumber\\
J_{ab} F_{bc} = F_{ab} J_{bc}.
\end{eqnarray}
The first of these means that $M$ is a holomorphic submanifold of $X$,
while the
second requires $F$ to be a $(1,1)$ form with respect to the complex
structure $M$ inherits from
$X$.  These conditions are familiar from both earlier worldsheet
analyses as well as the world-volume analysis of 
Becker {\it et al} \cite{BBS}.  
 \subsection{Differential Geometry Constraints}
\subsubsection{Constraint on the Spin Connection}

The first constraint we will address is Eq. (\ref{FERMSUSYCOND1}).
This is a condition on the curvature, $\omega_{azb} = \omega_{bza}.$  
We recall that $\omega_{azb} = e^I_a e^J_z \nabla_{I} e_{bJ} =
(e_z,\nabla_{e_a} e_b)$,
where $(\centerdot,\centerdot)$ denotes the Riemannian metric.  We
recall a basic result about the
Levi-Civita connection.  Since it is, by definition, symmetric, it
satisfies
$\nabla_X Y - \nabla_Y X = [X,Y]$ for any differentiable vector fields
$X,Y$.  So,
we see that the condition is equivalent to
\begin{equation}
(e_z,[e_a,e_b]) = 0.
\end{equation}
This is a necessary and sufficient condition that the collection of
vector spaces
$T_pM \subset T_pX$ for $p \in U\subset M$ spanned by the vector fields
$e_a$ contains
integral submanifolds!  If the brane wraps a submanifold of the target
space, then this condition is met.

\subsubsection{Constraints on $R$}
Now we will study the other two constraints, Eqs.
(\ref{FERMSUSYCOND2},\ref{SSUSYCOND1}).
Let us start with the latter.  Remembering that $T_{cf;e} = T_{cg}
H_{ghe} T_{hf},$ and $T^T R = T$,  we
write it as
\begin{equation}
\left[H_{ghe} + H_{egh} + H_{heg}\right] T_{gd} T_{hb} T_{ea} = 0.
\end{equation}
But, it is trivial to show that $H_{[abc]} = F_{[ab,c]}.$  Hence, SUSY
is implied by the Bianchi
identity for the $B$-gauge invariant field-strength $F_{ab}$ on the
brane.

Now let us show that Eq.(\ref{FERMSUSYCOND2}) is satisfied as well. 
Writing $T_{fd;e}$ as above, and using $T^T R = T$, we get
\begin{equation}
T_{ae} T_{ib} H_{ije} T_{jd} - 2 T_{ai} H_{ije} T_{jd} T_{eb} =
(b\leftrightarrow d).
\end{equation}
Since $T_{ae}$ is invertible, we get
\begin{equation}
\left(H_{ijk} -2 H_{kji}\right)T_{ib}T_{jd} = (b\leftrightarrow d).
\end{equation}
Using the antisymmetry of $H$ on the first two indices, we finally have
\begin{equation}
\left(H_{ijk} + 2 H_{jki} - H_{jik} - 2 H_{ikj}\right) T_{ib} T_{jd} =
0,
\end{equation}
which is exactly $H_{[ijk]} = 0$.  Again, the condition reduces to the
Bianchi identity
for $F$.  Since $F$ clearly satisfies the Bianchi identity, we conclude
that classically 
there are no constraints on the local geometry of supersymmetric cycles except
for the usual
co-isotropic/holomorphic conditions---Eqs. (\ref{COI},\ref{HOLO}).

\section{Discussion}
We have given a careful treatment of SUSY boundary conditions for
the $\Nc = (2,2)$ NLSM.  Starting from simple classical
mechanics notions we reproduced the well-known conditions on the
supersymmetric cycles.  Before we wrap up, we would like to reward
the reader's patience with a discussion of some interesting issues
raised in the text.
 \subsection{Locality constraint on the boundary action}
We have stressed that the fermion boundary couplings have no
effect on the supersymmetric cycles.  In fact, provided that
the boundary conditions are chosen to satisfy locality 
(Eqs. (\ref{FERMLOC},\ref{BOSB})), the fermion boundary 
coupling drops out of the action entirely.  However, in order
for locality to be satisfied, $C,D,\Ct$ must satisfy
\begin{equation}
\Ct = R D^T - D R^T - R^T C R,
\end{equation}
where we have used the SUSY condition $\Rt = R.$
A common fermion boundary coupling used in the literature is
\begin{equation}
F_{IJ} \left(\psi_+^I+\psi_-^I\right)\left(\psi_+^J+\psi_-^J\right).
\end{equation}
It was motivated by Callan {\it et al} as the exponentiation of the
open string photon vertex \cite{CLNY} operator, and more recently it
has been used in, for example, \cite{SW}.  This is a very natural
term to write.  Not only does it look like the exponentiation of the
the photon vertex operator, but in addition, by using the boundary
conditions it may be written as $4 (T^T F T)_{IJ} \psi_-^I\psi_-^J,$
a form that ensures that only the tangential components of $F$ enter
into the action.   Unfortunately, this natural term does not satisfy
the locality requirement.  Consider a space-filling brane.  Setting 
$\Ct = C = D = F,$ the locality constraint is written as
\begin{equation}
F(2 + R + R^T) = 0.
\end{equation}
It is easy to check that this is satisfied if and only if $F=0.$
So, although there are few constraints on the boundary fermion action,
the boundary actions in the literature do not seem to satisfy them!

Upon careful consideration, it is clear that there are subtleties 
associated with the ``exponentiation'' of the photon vertex.  Quite
simply, the naive exponentiation procedure does not take into account 
the change in the boundary conditions that accompanies turning on a
non-trivial $A(\phi)$ background.  This should be compared with the
exponentiation of closed string massless states into coherent state
backgrounds, where such subtleties do not arise.
 \subsection{Superfield Considerations}
In studying the SUSY of the boson boundary condition, we came
across a vexing problem:  the condition was SUSY, but only up
to the equations of motion!  This is fine for classical mechanics,
but it is certainly not satisfactory for quantum considerations.
How are we to remedy this?  There is one case where this is familiar 
to superspace aficionados: the supersymmetry algebra does not close 
off-shell once the auxiliary fields are integrated out of 
the theory.  Indeed, the algebra closes up to the equations of 
motion!  Here a similar situation holds, and the most optimistic 
way to interpret this is that there is a superspace formulation 
of this discussion where the boson boundary conditions are supersymmetric 
off-shell.  Thus, it would be very satisfactory to express this entire 
discussion in superspace.  By doing this, one might hope to obtain a 
description where the supersymmetry algebra closes off-shell and the 
boson boundary conditions are supersymmetric off-shell.  Even more 
fundamentally, one might hope that superspace would naturally provide 
a unique boundary action. 

\subsubsection{Adding auxiliary fields:  a toy example}
Let us try to include the auxiliary fields. To simplify matters, let us 
consider the case of flat target 
space and constant $F$.  Also, let us just worry about preserving an 
$\Nc = 1$ subalgebra of the manifest $\Nc=(1,1)$ SUSY.  
The $\Nc = (1,1)$ superfield has the component expansion
\begin{equation}
\Phi = \phi + i\theta^-\psi_- + i\theta^+\psi_+ + i \theta^-\theta^+ Y.
\end{equation}
If we demand that the fermions obey the boundary condition 
$\psi_+ = R\psi_-$ (we suppress target space indices for this discussion),
and that the boundary condition is supersymmetric under the $\ep$ SUSY,
we find that the bosons must satisfy
\begin{equation}
\label{TOYBOS}
\d_+\phi = R \d_-\phi - (1+R) Y.
\end{equation}
This is supersymmetric since the preserved supercharge squares to $i\d_0$.  
Furthermore, there is an elegant superspace expression for the boundary 
conditions:
\begin{equation}
\left[{\cal D}_+\Phi - R {\cal D}_-\Phi\right]_{\theta^+=\theta^-} = 0,
\end{equation}
where ${\cal D}_\pm$ are the superspace derivatives.  This looks nice,
but the rub is in trying to write down a sensible boundary action which will
produce the above as a locality constraint without spoiling the supersymmetry
of the action itself.  In fact, it is easy to convince oneself that this is
impossible.  A way out of this conundrum was suggested by 
Lindstr\"om {\it et al} in \cite{LRN}.  Since the local boson boundary 
conditions are $\d_+\phi = R \d_-\phi,$ it is natural to reconcile locality 
with Eq. (\ref{TOYBOS}) by introducing an additional boundary condition 
for the auxiliary field:
\begin{equation}
\label{TOYAUX}
(1+R) Y = 0.
\end{equation}
This is not quite enough.  We must also demand that $\delta_\ep \left[(1+R) Y\right] = 0.$
This leads to a final boundary condition:
\begin{equation}
\label{TOYEOM}
(1+R) \left( \d_-\psi_+ - \d_+\psi_-\right) = 0.
\end{equation}
These subsidiary boundary conditions can be written in superspace as
\begin{equation}
\left[(1+R) {\cal D}_+{\cal D}_- \Phi\right]_{\theta^+ = \theta^-} = 0.
\end{equation}
At first sight these look quite strange since they do not follow from
locality, and since the second one involves {\em derivatives} of the 
fermions on the boundary. One might worry that such a term spoils the 
initial value problem.  However, as pointed out in \cite{LRN}, these conditions 
are trivially satisfied on-shell since Eq. (\ref{TOYAUX}) is proportional to
the $Y$ equations of motion, and Eq. (\ref{TOYEOM}) is proportional to
the fermion equations of motion.  It is easy to check explicitly 
that these boundary conditions are supersymmetric.  

Thus, at least in this toy example, we can find a supersymmetric version of 
the boundary conditions.  The price to pay is the (expected) introduction 
of auxiliary fields $Y$ for the closure of the supersymmetry algebra and 
additional boundary conditions---Eqs. (\ref{TOYAUX},\ref{TOYEOM}) which 
do not follow from locality.

\subsubsection{A superspace action}
The toy example above suggests that it is possible to introduce auxiliary fields and
supplementary boundary conditions so that the action and the boundary conditions are
supersymmetric off-shell.  To study the more general curved background, it is convenient
to work in superspace.  Consider once again the bulk $\Nc=(1,1)$ NLSM (see, for example, 
\cite{JOPO} for superspace conventions). In the presence of a
boundary the action transforms under the diagonal $\Nc=1$ supersymmetry
generated by the unbroken supercharge ${\cal Q} = {\cal Q}_+ + {\cal Q}_-$.  
Of course, one can choose 
boundary conditions that ensure the invariance of the action.  Alternatively, 
it is fairly straightforward to add a boundary Lagrangian to make the action $B$-gauge
invariant and of the form 
\begin{equation}
S = \int_{\Sigma} \left[{\cal Q}, \centerdot\right] + \int_{\d\Sigma} \left[{\cal Q},\centerdot\right],
\end{equation}
where $\left[{\cal Q},O\right]$ denotes the SUSY action on $O$.
The ``improved'' total action is given by 
\begin{align}
\label{IMPRSUPERACTION}
S = 
& \
-\frac{1}{4} \int_\Sigma d^2 x \ {\cal D} \tilde{\cal D} 	
		\left.
			\left[ 
				\left( G_{IJ}(\Phi) + B_{IJ}(\Phi)\right) {\cal D}_+\Phi^I {\cal D}_-\Phi^J 
			\right]
		\right|_{\theta^\pm = 0} 
\nonumber \\ 
&
    	- \frac{i}{2} \int_{\partial\Sigma} d x^0 \ {\cal D}
    		\left.	
			\left[
				A_I(\Phi) {\cal D} \Phi^I 
			\right]
		\right|_{\theta^\pm = 0}.
\end{align}
Here ${\cal D} = {\cal D}_+ + {\cal D}_-$
and $\tilde{\cal D} = {\cal D}_+ - {\cal D}_-$
correspond to the preserved and to the broken linear combinations 
of supercharges respectively.\footnote{Note that in superspace the algebraic statement 
$\left[{\cal Q},O \right]$ is geometrized into 
${\cal D}\left(O \right).$}  
Since ${\cal Q}^2 = 2 i\d_0$, this action is obviously SUSY
invariant.  We can consider the variation of action under 
the $\ep'$ SUSY as well.  One can show that the total action is invariant {\em without}
the use of boundary conditions under the $\ep'$ variation for {\bf B}-type SUSY , but 
not for {\bf A}-type SUSY, where boundary conditions are needed for invariance.  This is 
to be expected: {\bf B}-type SUSY is compatible with the holomorphic structure of $\Nc=(2,2)$
superspace, while {\bf A}-type SUSY requires reality conditions that are, in a sense, ``unnatural'' 
from the holomorphic $\Nc = (2,2)$ point of view.\footnote{One can add a different boundary
term that ensures invariance of the action under {\bf A}-type SUSY without the use of
boundary conditions.  In components, it is of the form 
$G_{IJ}Y^I\phi^J + \frac{1}{2} \d_1(\phi^I G_{IJ}\phi^J)$.  Unfortunately, this term is incompatible
with locality.}

To investigate how locality is affected by the addition of this improvement
term, we write the variation of the action as
\begin{align}
\label{ACTVARIATION}
\delta S = 
&
	\int_\Sigma d^2 x\ {\cal D}_+ {\cal D}_- 	
		\left.
			\left[ 
				\left( 
					G_{IJ} {\cal D}_-{\cal D}_+\Phi^I 
					+ G_{IJ}  \Gamma^{I}_{~LM} {\cal D}_-\Phi^L {\cal D}_+\Phi^M
				\right) \delta \Phi^J 
			\right]
		\right|_{\theta^\pm = 0}
\nonumber \\ 
&
    	+ \frac{i}{2} \int_{\partial\Sigma} d x^0 \  {\cal D}
     		\left.
 			\left[
 				\left( 
 		F_{IJ} {\cal D}\Phi^I + G_{IJ} \tilde{\cal D}\Phi^I
 				\right) \delta \Phi^J 
 			\right]
 		\right|_{\theta^\pm = 0}
\\ 
&
     	- i \int_{\partial\Sigma} d x^0 \ 
     		\left.
 			\left[
 				 \left( 
 					G_{IJ} {\cal D}_-{\cal D}_+\Phi^I 
					+ G_{IJ} \Gamma^{I}_{~LM} {\cal D}_-\Phi^L {\cal D}_+\Phi^M
 				\right) \delta \Phi^J 
 			\right]
		\right|_{\theta^\pm = 0}.
\nonumber
\end{align}
The first line encodes the bulk equations of motion for the fields.  The second
and third lines encode the boundary terms in the variation that need to be
eliminated by an appropriate choice of boundary conditions.  The term in the
second line are explicitly supersymmetric, since they are expressed as 
$\left[{\cal Q},\centerdot\right]$.
The third line precisely contains the non-supersymmetric contribution to
the boundary conditions.  One can show that these contributions will vanish
if we impose the following subsidiary boundary conditions in addition to locality:
\begin{equation}
\label{SUPERY}
T^I_{~J} \left( Y^J - i \Gamma^J_{~KL} \psi_-^K\psi_+^L\right) = 0,
\end{equation}
and 
\begin{equation}
\label{SUPEREOM}
T^{IJ} \left(E^+_J - E^-_J\right) = 0,
\end{equation}
where $E^\pm$ are the fermion equations of motion (Eq. (\ref{EOM})).
As in the toy example, the two additional boundary conditions are trivially
satisfied on-shell:  Eq. (\ref{SUPERY}) follows from the $Y$ equations of
motion, while Eq. (\ref{SUPEREOM}) follows from the fermion equations of
motion.  

The subsidiary boundary conditions can be imposed dynamically by introducing
boundary superfield Lagrange multipliers,  
$\Lambda^I (x^0) = \lambda^I (x^0) + \theta l^I (x^0)$ into the action:
\begin{equation}
\int_{\partial\Sigma} d x^0 \ {\cal D}
        \left.  
                \left[
                        \Lambda^I       \left( 
                                        G_{IJ} {\cal D}_-{\cal D}_+\Phi^J 
                                        + G_{IJ} \Gamma^{J}_{~LM} {\cal D}_-\Phi^L {\cal D}_+\Phi^M
                                        \right) 
                \right]
        \right|_{\theta^\pm = 0}.
\end{equation}
$\Lambda^I$ is a fermionic superfield with non-zero components in the directions tangent 
to the cycle.

As promised above, the superspace approach also suggests a natural form for 
the fermion bilinear coupling on the boundary.  It is
\begin{equation}
-\frac{i}{4}\left[F_{IJ} \left(\psi_+^I+\psi_-^I\right)\left(\psi_+^J+\psi_-^J\right) 
                         + 2G_{IJ}\psi_+^I\psi_-^J\right],
\end{equation}
which automatically satisfies the locality condition (Eq.  (\ref{FERMLOC})).
Though this form is not not uniquely determined by requirements
of locality, it arises so naturally from superspace that it would not be
too surprising if this is precisely what one would obtain by a careful
exponentiation of the photon vertex operator.  This action has also been
obtained by a different method in \cite{HASSAN}.
%
\section{Acknowledgments}
 We would like to thank P.S. Aspinwall, R. Bryant, C. Curto, D. Fox, A.
Kapustin, R. Karp, D.R. Morrison, K. Narayan, M. Ro\v{c}ek 
and M. Stern for useful discussions. One of us (IVM) would like to
thank the organizers of
TASI 2003, where some of this work was completed. 
MRP and SR thank KITP and the organisers of
the Geometry, Topology, and Strings workshop (MP03) where some of this
work was completed.  Their participation was supported in part by  the
National Science Foundation under Grant No. PHY99-07949.  
This work is partially supported by NSF grant DMS-0074072.
\bibliographystyle{unsrt}

\end{document}